\documentstyle[11pt,newpasp,twoside,epsfig]{article}

\begin{document}
\def\cd{cd$^{-1}$}
\def\cds{cd$^{-1}$\,}
\def\kms{km~s$^{-1}$}

\title{Mode detection from line-profile variations}
 
\author{L. Mantegazza}
\affil{Osservatorio Astronomico di Brera, Merate, Italy
luciano@merate.mi.astro.it}




\begin{abstract}
The phenomenology associated with line profile variations in
$\delta$ Scuti stars is reviewed. The three main techniques
adopted to detect pulsation modes, i.e., the pixel--by--pixel analysis, 
the analysis of the moment time series and the Fourier Doppler Imaging
are presented and discussed. Their application to the observational data 
has allowed the reliable detection of several pulsation modes, many of them
not detectable with photometric observations. The need of better 
coordination of simultaneous spectroscopic and photometric campaigns 
in order to get the complete pulsation spectra of these stars is pointed
out. The observation in different campaigns of the same objects has
proved that dramatic changes in the mode amplitudes are quite common,
even in a couple of years (and maybe less).

In the last part of the paper a recent approach to the mode detection
by direct fit of line profile variations in multiperiodic pulsators
is described.
\end{abstract}




\section{Historical development}

While the presence of radial velocity variations in $\delta$ Scuti itself
was detected just at the beginning of the century (Wright 1900), the
first hints about the possible presence in the same star of line profile
variations (hereinafter LPV) were due to Struve about half a century later
(Struve 1953).

After a decade of quiescence the studies of the phenomena associated
with LPV  were resumed by several authors in the mid sixties and
seventies. Several different phenomena were announced, such as variable
emission in the cores of strong lines, large equivalent width variations
and so on (for a review see the paper by Breger 1979).  Unfortunately
those findings were probably connected to the use of the photographic
plate as a detector. As a matter of fact later studies of the same objects
performed with photoelectric detectors were not able to confirm any of
those findings.

The first reliable observations of LPV in $\delta$ Scuti stars  were
performed at the beginning of the eighties with the advent of the
Reticon detector. Several stars were observed by Campos \& Smith (1980)
and Smith (1982) but unluckily most of them for a few hours in one night
only, i.e., while the presence of line profile variations was clearly
detected, the data were not sufficient to analyze the variations and to
detect periodicities. Generally the few data were phased with the periods
available from photometry or radial velocity curves (usually at that time
only one or two periods per star were known).  Among the stars studied by
these authors only 28 Aql had a sufficient dataset (4 consecutive nights)
which allowed the analysis of line width and radial velocity variations.
A period--finding program confirmed the presence in both parameters of
the two known periods derived from photometry  (Smith 1982).

The successive papers showed, beside the changing shape of the line
profiles, the presence of moving sub-features in the lines (e.g., Yang \&
Walker 1986, Walker et al. 1987).  In these cases the data were also not
adequate for independent period searches, although two attempts were made
by Kennelly et al. (1991, 1992), who analyzed 2.5 hours and 5 hours of
observations of $k^2$ Boo and $\gamma$ Boo, respectively. They did it
by sampling at 10 km~s$^{-1}$ intervals the intensity residuals in the
line profiles derived from the subtraction of the average profile, and
computing the Discrete Fourier Transform. Obviously, given the extremely
short temporal baseline, the frequency resolution was very poor, allowing
only a rough confirmation of the photometric periods.

In the following years it was realized that more intensive campaigns on
a few selected objects were more fruitful than sporadic observations of
many targets.  This was in part the consequence of the photometric studies
that were showing that usually many close frequencies were present in
the pulsation spectra and consequently longer baselines were needed in
order to resolve these frequencies.

Now before discussing in detail the results concerning these campaigns
we shall briefly review the phenomenology associated with the LPV,
the constraints that it imposes on the observational programmes and the
techniques we can adopt to analyze the data.
  
\section{Phenomenology}

Fig. 1 shows the LPV observed during one night in the spectrum of BV
Cir; in order to clearly show the LPV  the average spectrum has been
subtracted from each spectrum. We see the presence of waves propagating
across the profiles of the  lines at 4501, 4508 and 4515 \AA~from short
to long wavelengths, while in the continuum regions there are only noise
fluctuations. In the figure the borders of the three lines have been
marked by the solid vertical lines.
\begin{figure}[ht]
\setlength{\textwidth}{4.3in}
\plotone{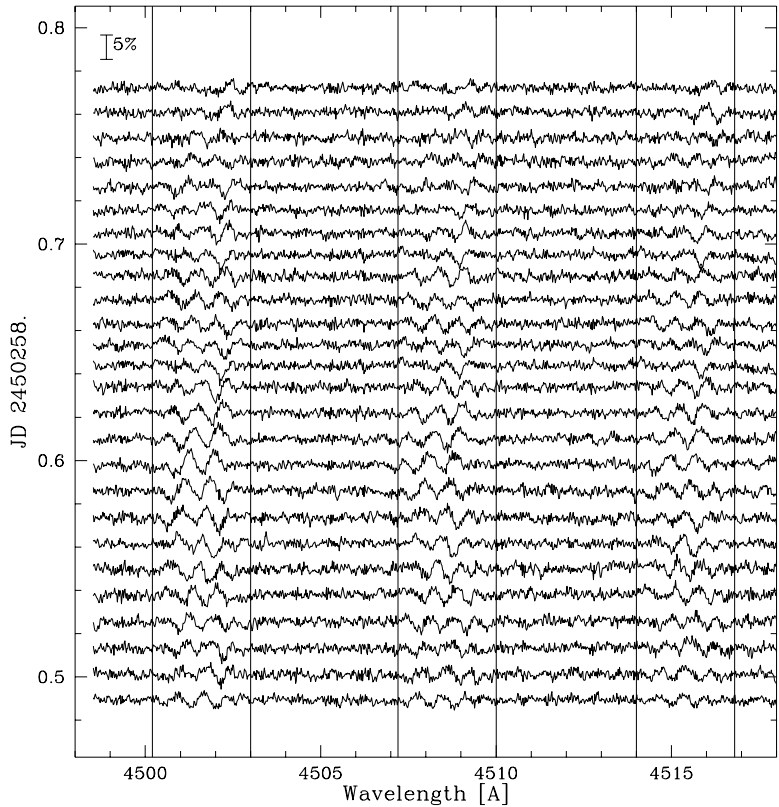}
\caption{Variations in the line profiles of the lines 4501, 4508 and
4515 \AA~during one night of observations. The average spectrum has
been subtracted to the individual ones in order to better evidence the
variations. The small bar at the upper left indicates an amplitude of  5\%
in the continuum intensity.  The vertical solid lines individuate the
borders of the three lines.  The fluctuations in the continuum regions
allow to estimate the $S/N$ which in this case is about 250.
\label{fig-1}
}
\end{figure} 

As we can judge from the figure the typical time-scale for a feature
to cross the line is of the order of a few hours, while the amplitudes
of these features are of the order of a few thousandths of the continuum
intensity.

As is well known we can interpret such variations as the result of
the presence of non-radial pulsations: in each zone of the visible
stellar disk the pulsation velocity combines with the rotational one and
therefore the flux from this zone, because of the Doppler effect, gives
a contribution to the line profile at a distance from the line center
corresponding to its radial velocity. In this sense, if the pulsation
velocity is small with respect to the rotational one,  the line profile
broadened by rotation supplies a one-dimensional ``Doppler imaging''
of the stellar disk (Vogt \& Penrod 1983).  For this reason it is a
common practice, when describing the variations across the line profiles,
to transform wavelengths into velocities by means of the Doppler formula,
assuming the zero velocity at the line center (see, e.g., Fig. 10).

In the case of Fig. 1 the moving features indicate that the prevailing
pulsation modes are crossing the visible stellar disk moving in the same
direction of rotation (i.e., they are prograde as seen by the observer).

If the set of spectra has a rather good phase coverage of the periods
of the different pulsation modes we can get a first guess of some
quantities that we need in the further data analysis.  First, we can
compute the average of the spectra and assume that this approximates
the nonpulsation spectrum of the star (however it can be seen that this
average spectrum tends to have broader lines that the true nonpulsation
one, see, e.g., Hao 1998).  Then we can estimate at each wavelength step
(or for each pixel of the spectrum) the standard deviation of the
normalized flux about the mean.
\begin{figure}[ht]
\setlength{\textwidth}{4.3in}
\plotone{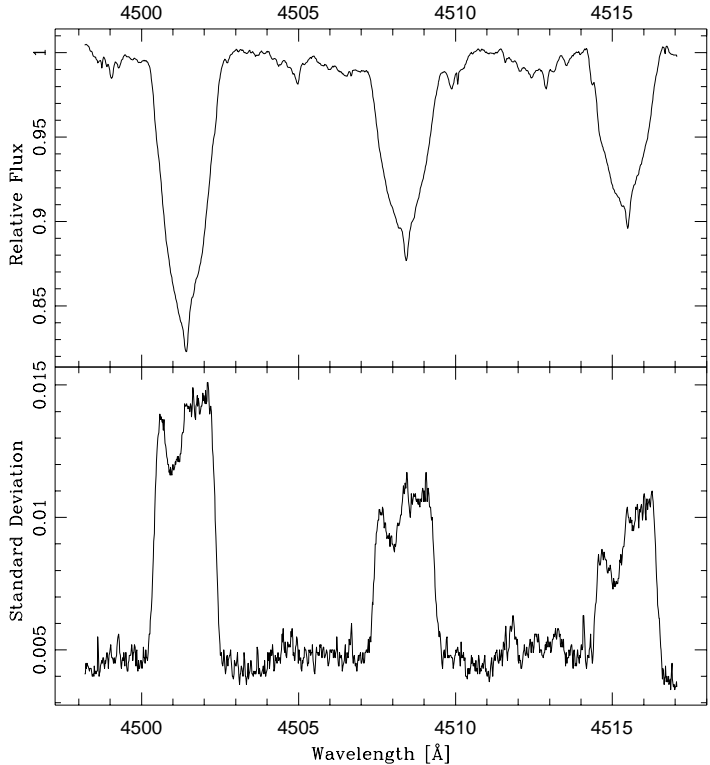}
\caption{Average spectrum of X Caeli from the data obtained in the 1996
campaign (top panel); standard deviation about the mean of the flux
measured in each pixel (bottom panel).
\label{fig-2}
}
\end{figure} 
\begin{figure}[ht]
\setlength{\textwidth}{4.3in}
\plotone{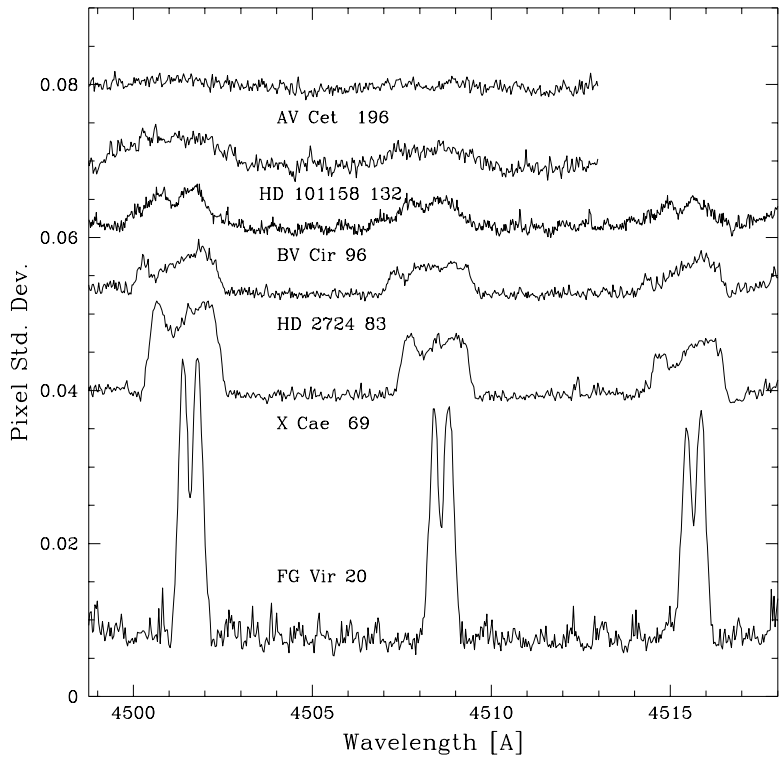}
\caption{Standard deviations of the normalized flux in each pixel in the
same spectral region of six $\delta$ Scuti stars. The number beside each
star's name is its projected rotational velocity in km/s. The spectra have
been vertically shifted to avoid superposition, but the scale is the same.
\label{fig-3}
}
\end{figure} 

As an example Fig. 2 shows in the upper panel the average spectrum of
X Caeli while the lower panel shows the standard deviation of the individual
spectra about the mean. This panel clearly shows that the
variability is restricted inside the line profiles, furthermore it
allows the definition  for each line of the limits in which the profile
variations are present. In the continuum zones the reciprocal of the
standard deviation of the normalized fluxes about the continuum supplies
an estimate of the $S/N$ of each spectrum, and the reciprocal of the
average of the values of the standard deviations of all the spectra about
the average one in the same zones gives an estimate of the average $S/N$
at the continuum level. For example in the X Cae case the lower panel
of Fig. 2 suggests an average $S/N$ of about 230\footnote{As $S/N$
in Astronomy is usually intended the ratio between amplitudes and not
between powers as preferred by engineers.}.

Fig. 3 shows the standard deviation of the normalized flux in each
pixel for six stars arranged, from the bottom to the top, in order of
increasing projected rotational velocity. We can see that as the line
profiles broaden the flux variation decreases, until for the fastest
rotator ($v\sin i=196$~\kms) this variation is barely discernible.
It can be also appreciated that it becomes increasingly difficult, as
$v\sin i $ increases, to find unblended lines and continuum regions
which can be used for the spectrum normalization.

While it is obvious that in order to detect high-degree modes it is
necessary to study stars with high $v\sin i$ \footnote{As a rule of
thumb the highest detectable degree is of the order of $v\sin i/W_i$
(Kennelly et al. 1992), where $W_i$ is the intrinsic line width, i.e., the
width that the line should have if rotational broadening was not present.}
, if we consider low-degree modes, it can be shown that in weak lines
their $S/N$ ratio decreases approximately with $(v\sin i)^{1/2}$,
and therefore these modes are more easily detectable in stars with low
projected rotational velocity.

Starting from the average line profile we can also get a first estimate
of the projected rotational velocity and of the intrinsic line width.
\begin{figure}[ht]
\setlength{\textwidth}{3.0in}
\plotone{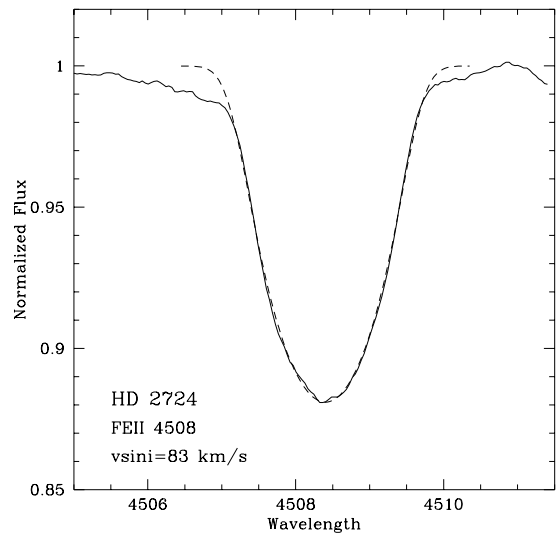}
\caption{Average profile of the FeII 4508\AA~line of the star HD2724
(BB Phe) as observed in the 1997 campaign (solid line); rotationally
broadened intrinsic profile that best-fit the average one (dashed line). A
projected rotational velocity of 83~\kms has been derived.
\label{fig-4}
}
\end{figure} 
This can be accomplished by performing a non-linear least-squares fit of a
rotationally broadened Gaussian profile to the observed one.  Fig. 4 shows
as an example the fit derived in this way on the FeII 4508 \AA~line of
the star HD 2724 (Mantegazza \& Poretti 1998).  These two quantities
can also be estimated using the zero-points of second and fourth moment
(Balona 1986b)\footnote{For the definition of the line moments see the
paper by Aerts \& Eyer, these proceedings.}.  As we have previously said,
the average profile tends to be larger than the nonpulsation one. A way
to get a better estimate of these two quantities is discussed at the
end of this paper.

\section{Observational constraints}

From what we have shown in the previous section we can estimate what
the observational constraints should be in order to detect and to study
the LPV in $\delta$ Scuti stars.

The typical time-scales of the observed variations (a few hours) impose
a limit on the integration times, which should be a fraction of these
quantities.  According to our experience this limitation is not very
stringent for the mode detection, for instance in the case of FG Vir
we have detected the 34.1 \cd~ mode with an integration time of 15 min,
which corresponds to about 35\% of the pulsation period (see Table 1).

For the mode identification the discourse is more delicate, because even
if with a relatively long integration time we are still able to detect
the fundamental frequency of a mode and to correct its amplitude for
the damping due to the integration in time, its harmonics, which are
of lesser amplitudes, can be lost in the noise, because of the larger
damping due to their shorter periods, and therefore we would not be able
to recover the correct shape of the signal.

The amplitudes of the perturbation across the line profiles are for most
of the modes of the order of a few thousandths of the continuum intensity,
and therefore, to be reliably detected and measured, we need a $S/N$
of at least a few hundred.

Finally, it remains to be evaluated what the minimum spectral
resolution should be. Since, as it is explained in the paper by Aerts \&
Eyer (this proceedings), we can approximate the observed line profiles as
the convolution between an intrinsic profile ($W_i$) and the rotational
broadening one, the resolution should be 
$$R=\lambda/\delta\lambda\geq c/W_i .$$
Since typically $W_i\simeq10 km/s$, then $R\geq 30000$.

The three requirements: short exposures, high $S/N$ and high resolution,
impose that we collect a lot of photons in short times, and this implies
that either we limit ourselves to the study of very bright targets,
or we observe with large telescopes, or, by observing a wide spectral
region, we find a way to add the information of the LPV contained in
many of the observed lines without blurring it.

Up to now 4 different approaches have been adopted to add the information
of several lines: a) the straight average of the individual line profiles
(e.g., Kennelly et al. 1996); b) the average of the line moments (e.g.,
Mantegazza \& Poretti 1996), c) the computation of the cross-correlation
function (e.g., Korzennik et al. 1995), d) the deconvolution of the
observed profiles by the intrinsic one (Kennelly et al. 1998, see also the
paper by Aerts \& Eyer, these proceedings). None of these approaches are 
free of problems, and in particular all of them assume that the intrinsic
profile should be the same for all the lines and therefore in all cases
a careful selection of lines with similar characteristics should be done.

\begin{figure}[ht]
\setlength{\textwidth}{4.3in}
\plotone{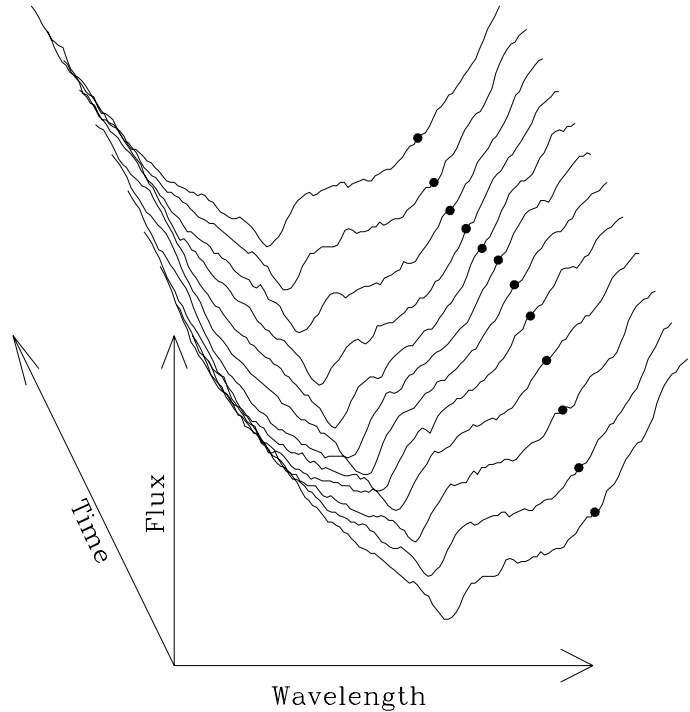}
\caption{Line profile variations of the 4501\AA~line in the star X Caeli
phased according to its dominant period (7.39\cds). The dots indicate
the flux at a fixed pixel in the line profile.
\label{fig-5}
}
\end{figure} 

\section{Analysis techniques}

Now we shall briefly review how we can study the variations in a line 
profile.
Three different techniques have been adopted to search for the periodicities
present in LPV:
\begin{itemize}
\item[a)] the pixel-by-pixel analysis;
\item[b)] the frequency analysis of the moment time series;
\item[c)] Fourier Doppler imaging.
\end{itemize}

\subsection{Pixel-by-pixel}

The pixel-by-pixel analysis is based on the fact that during an
oscillation cycle the flux measured in the same pixel of the line profile
(i.e., at the same wavelength) fluctuates with the same period. Fig. 5
shows this fact for the TiII 4501 line of the star X Caeli. The profiles
have been phased according to the dominant pulsation mode (7.39 \cd)
and cover one complete cycle. The dot indicates for each profile the flux at
a given fixed wavelength.  Therefore we can extract from the set of all the
profiles of the same line the time series constituted by the fluxes in the
same pixel and then analyze them with the usual techniques developed to
study one--dimensional time series (e.g., light-curves). We have seen in
the previous section how the diagram showing the flux standard deviation
in each pixel can be used to define the line borders and hence to detect
which pixels can supply useful time series.

Before extracting the individual pixel time series it is necessary
to rebin the spectrograms in order to remove the observer's
velocity variations due to the Earth's revolution and rotation. Even
the latter cannot be neglected because its amplitude (about 0.5~\kms)
can be comparable to those of the pulsation modes.

Two techniques have been the most frequently used to analyze the
pixel-by-pixel time series: a) the least-squares power spectrum
(Vani\^cek 1971), b) the CLEAN algorithm (Roberts et al. 1987)
\begin{figure}[ht]
\setlength{\textwidth}{4.3in}
\plotone{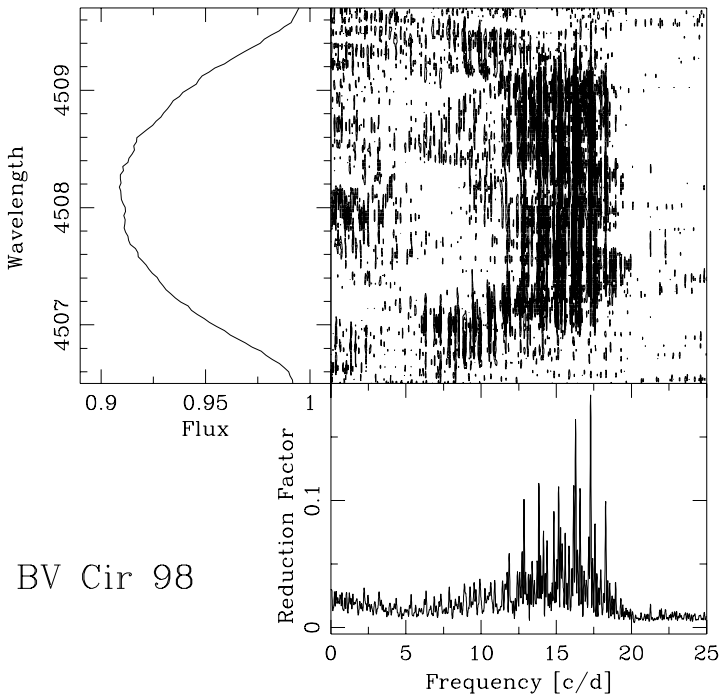}
\caption{Frequency analysis of the variations in the profile of the
FeII 4508 \AA ~line of the star BV Cir.  Left panel: Average profile.
Central panel: pixel-by-pixel least--squares power spectra without known
constituents (see text). Bottom panel ``global'' least-squares power
spectrum (see text).
\label{fig-6}
}
\end{figure} 

Fig. 6 shows the application of this approach to analyze the LPV in
the FeII 4508 line of BV Cir (1998 campaign, Mantegazza  et al. 1999).
The upper left panel shows the average line profile, the central panel
contains the pixel-by-pixel spectra computed with the least--squares
technique without known constituents (i.e., the spectrum contains all
signals present in the line variations, for details see below), and the
lower panel shows the global least-squares spectrum (which is equivalent
to a weighted average of the individual pixel-by-pixel spectra).
In this case we are in the presence of rather complicated LPV which contain
several modes with low, medium and high frequencies in the observer's
reference frame\footnote{Because of stellar rotation a non-axisymmetric non-radial mode
with an oscillation frequency $\nu_0$ in a reference frame co-rotating
with the star is seen from the observer as having a frequency $\nu=\nu_0-m
\Omega$, where $m$ is the mode azimuthal order and $\Omega$ is the
stellar rotational frequency.}.
Since the data were obtained in a single site campaign the peak of each
mode is flanked by rather strong 1 \cd aliases, and this, coupled with
the intrinsic complexity of the pulsation spectrum, explains why the
computed power spectrum is so entangled. In order to have a clear picture
of the true pulsation spectrum is necessary to detect one by one all the
periodicities or to analyze the data with the CLEAN algorithm.  The result
of the application of this technique is shown in Fig. 7.  Other examples
of the application of the CLEAN algorithm to the pixel-by-pixel analysis
are in the papers by Bossi et al. (1998) and De Mey et al. (1998).

\begin{figure}[ht]
\setlength{\textwidth}{4.3in}
\plotone{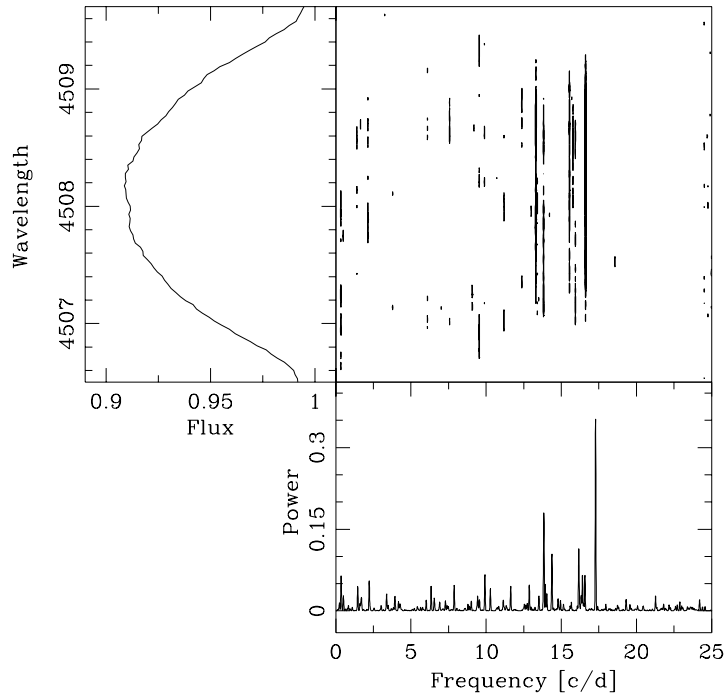}
\caption{Frequency analysis of the variations in the profile of the FeII
4508\AA~line of the star BV Cir.Left panel: Average profile. Central
panel pixel--by--pixel CLEAN power spectra. Bottom panel: average of
the pixel--by--pixel CLEAN spectra.
\label{fig-7}}
\end{figure} 

The power spectrum of Fig. 7 is much more easily readable than that of
Fig. 6 even if there are some aliases still present and probably in some
cases aliases have been preferred to the true peaks. The reason is due
to the fact that the CLEAN approach to select the peaks is completely
automatic, and among other things it does not allow to select between
competing aliases the use of the a priori knowledge, such as  for example
the fact that we know some correct periods from the light-curve analysis.

In our opinion and according to our experience, the most reliable approach
for looking for multi-periodicities is to proceed to detect one by one
the periods with the least squares technique. This approach has been
slightly modified to best analyze LPV (Mantegazza \& Poretti 1999) in
order to define a ``global'' least-squares spectrum which includes the
information of the variability of the whole profile.

\subsubsection{The least-squares algorithm generalized to study LPV}

In general our data consist of a set of profiles of a spectral line
$P(\lambda_j,t_k)$ ($j$ is the pixel number and $t_k$ is the time of
the $k$-th spectrogram) whose global variance can be defined as

$$\sigma_T=\sum_{j,k}w_k^2\left[P(\lambda_j,t_k)-P_0(\lambda_j)\right]^2 \eqno(1)$$ 
where $P_0(\lambda_j)$ is the time averaged profile and the $w_k$'s are the
normalized weights derived from the $S/N$ of the individual spectrograms.

The frequency analysis of the variations present in these profiles
can be iteratively performed in the following way: if we have already
detected $m$ periodic sinusoidal components (``known constituents'')
and we are looking for the $m+1$, we explore the useful frequency range
(0$<\nu_i<\nu_{max}$ \cd) by fitting each pixel time series $j$ with
the series
\setcounter{equation}{1}
\begin{eqnarray}
 p_{i,j}(t_k)=\overline p_{i}+\sum_{l=1,m} A_{i,j,l} \cos(2\pi\nu_l t_k
+\phi_{i,j,l}) \nonumber \\
+ A_{i,j,m+1} \cos(2\pi\nu_i t_k+\phi_{i,j,m+1}) 
\end{eqnarray}
where $\overline p_{i},A_{i,j,l},\phi_{i,j,l}$ (with $1\leq l\leq m+1$)
 are the free parameters. 

Then we compute the global reduction of variance defined as

$$RF_i=1-\sum_{j,k}w_k^2(p_{i,j}(t_k)-P(\lambda_j,t_k))^2/\sigma_m \eqno(3)$$ 
where $\sigma_m$ is the global residual variance after the fit of the line
profile variations with the $m$ ``known constituents''.  The frequency
$\nu_i$ giving the highest $RF$ (or one of its 1~\cds aliases if
there is a better agreement with the photometrically detected modes)
is then selected as the $m+1$--th known constituent and the procedure
is iterated again.

The search ends when no dominant peaks are apparent in the last
spectrum. On this point a more quantitative criterion should be developed
such as the one adopted by Breger et al. (1995) to assess the
physical reality of peaks detected from light-curve frequency analysis.

At the end of this procedure, after having detected M known constituents,
we can perform a final fit of $P(\lambda_j,t_k)$ and derive the functions:
$\overline p_M(\lambda_j)$ (i.e., the best estimate of the average line
profile), $A_l(\lambda_j), \phi_l(\lambda_j)$ (with $1<l<M$) and also
their formal errors.

As an example we report in Fig. 8 the frequency analysis of the variations
in the 4508\AA~ line of the star HD2724. Each panel contains a ``global
least-squares spectrum'' computed by introducing the frequencies of the
terms detected in the previous spectra as ``known constituents''.
\begin{figure}
\setlength{\textwidth}{4.3in}
\plotone{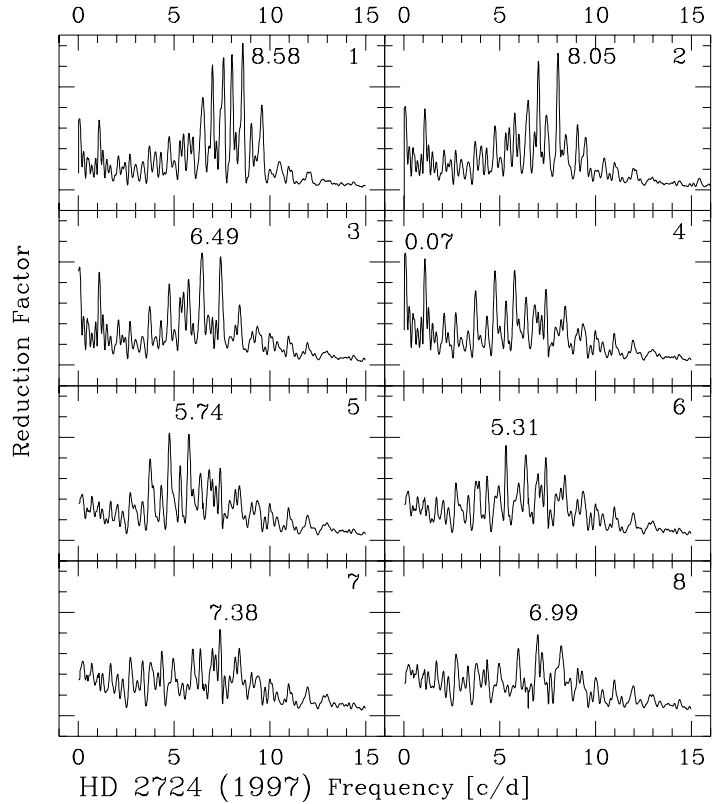}
\caption{Complete frequency analysis with the global least-squares power
spectra of the variations in the profile of the 4508\AA~ line of HD
2724 (1997 campaign).  In each successive spectrum all the frequencies
detected in the previous ones are given as ``known constituents''. The
order of computation follows the numbering reported on the upper left
of each panel. In case of ambiguity between 1 c/d aliases the choice of
the correct peak has been guided by the knowledge of the more accurate
photometrically detected frequencies. The selected frequency at each
step of the analysis is written beside the respective peak.
\label{fig-8}}
\end{figure} 
Fig. 9 summarizes in the upper panel the pulsation spectrum as derived
from the frequency analysis of the previous figure. In the lower panel
we show the same spectrum as obtained using the CLEAN algorithm. In
this case the two approaches supply the same mode detection (the only
discrepancy is on the selection of the lowest frequency peak where there
is an uncertainty of 1 c/d, but as we can see in panel 4 of Fig. 8,
the difference between the two peaks is very marginal).

\begin{figure}
\setlength{\textwidth}{4.3in}
\plotone{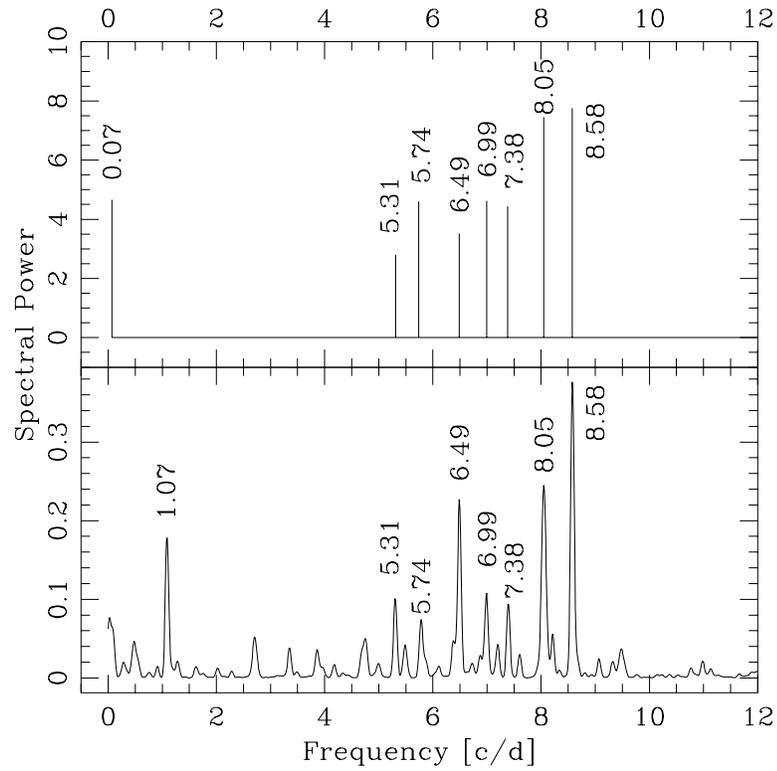}
\caption{Upper panel: pulsation spectrum of HD 2724 as derived from the
least-squares power spectrum analysis shown in the previous figure. In
the ordinates the squared amplitudes averaged across the whole profile
are shown. Lower panel: pixel-by-pixel CLEAN spectrum of the same data.
\label{fig-9}}
\end{figure} 

Fig. 10 shows the behaviors of amplitudes and phases across the 4508 \AA~
line profile of the terms detected in the previous example (functions
$A_l(\lambda_j), \phi_l(\lambda_j)$ and their corresponding formal
errors).  The positions on the line profile are expressed in Doppler
velocities as described in Section 2.  The term at 0.07 \cd~ is not
reported because it was probably introduced by an instrumental effect
(see Mantegazza \& Poretti 1999).  It can be observed that the phase
diagram of the 5.31 \cd~ term has the typical behavior of a low--degree
retrograde mode (its phase increases in the same direction as the stellar
rotation, i.e., with the wavelength). The opposite behavior is shown
by the phases of the 8.58 \cd~ term, which is typical of a moderately
high--degree prograde mode.
\begin{figure}
\setlength{\textwidth}{4.3in}
\plotone{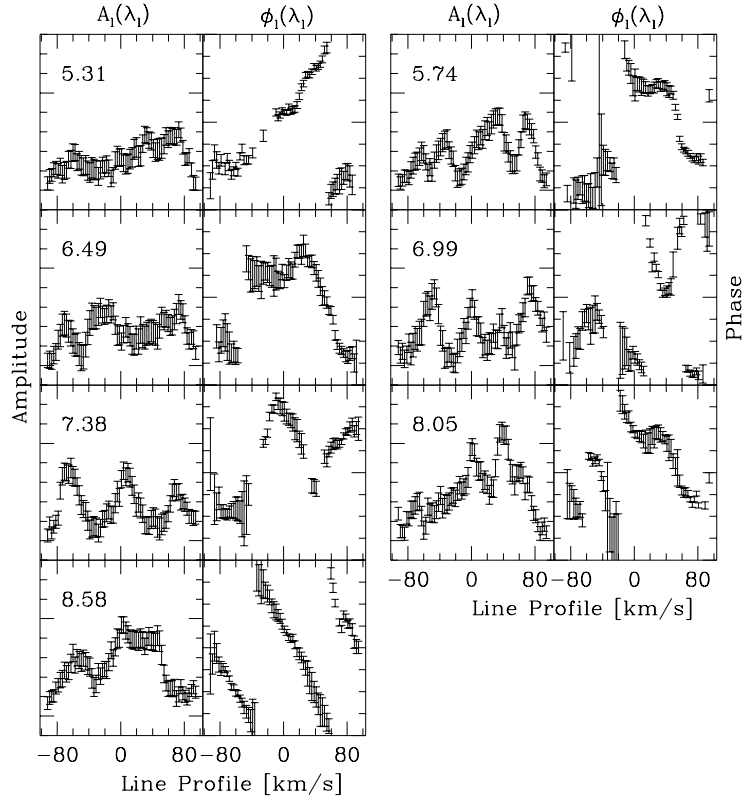}
\caption{Behavior of amplitudes (first and third columns) and phases
(second and fourth columns) across the profile of the 4508\AA~line
of the modes detected in the frequency analysis reported in Fig. 8
(functions $A_l(\lambda_j), \phi_l(\lambda_j)$ of the text). The bars
are the formal errors derived as described in the text.  All the diagrams
have the same scale: amplitudes are in units of the continuum intensity
(each tick corresponds to 0.001) and phases are in degrees (each tick
corresponds to $50\deg$). The ranges  of the diagrams are from 0 to 0.008
and from -200$\deg$ to +200$\deg$ for amplitudes and phases, respectively.
The frequency (in \cd) of each mode is reported at the upper left of the panel
containing the respective amplitude function.
\label{fig-10}}
\end{figure} 

The diagrams that show the behavior of the phases of each mode across the
line profile are very instructive to show the nature of the non-radial
modes present in $\delta$ Scuti stars.  In order to demonstrate this
Fig. 11 shows a zoo of behaviors of such phases as we have derived
them from our studies. Seven different modes are represented in the
panels ranging from a high-degree retrograde mode (bottom panel) to
a high-degree prograde mode and passing from an axisymmetric mode ($m=0$, 
third panel from the bottom).  For each panel we give the name
of the star to which the mode belongs, the frequency of the mode in
the observer's reference frame and its $\ell$ and $m$, as derived from
the LPV fit technique described in the last section of this paper.
The abscissae give the positions across the line profile with Doppler
velocities (or equivalently wavelengths) increasing from left to right.
+1 and --1 indicate a distance from the center corresponding to $+v\sin
i$ and $-v\sin i$ respectively (i.e., in the Doppler map they correspond
approximately to the limbs of the visible stellar disk).
\begin{figure}[ht]
\setlength{\textwidth}{4.3in}
\plotone{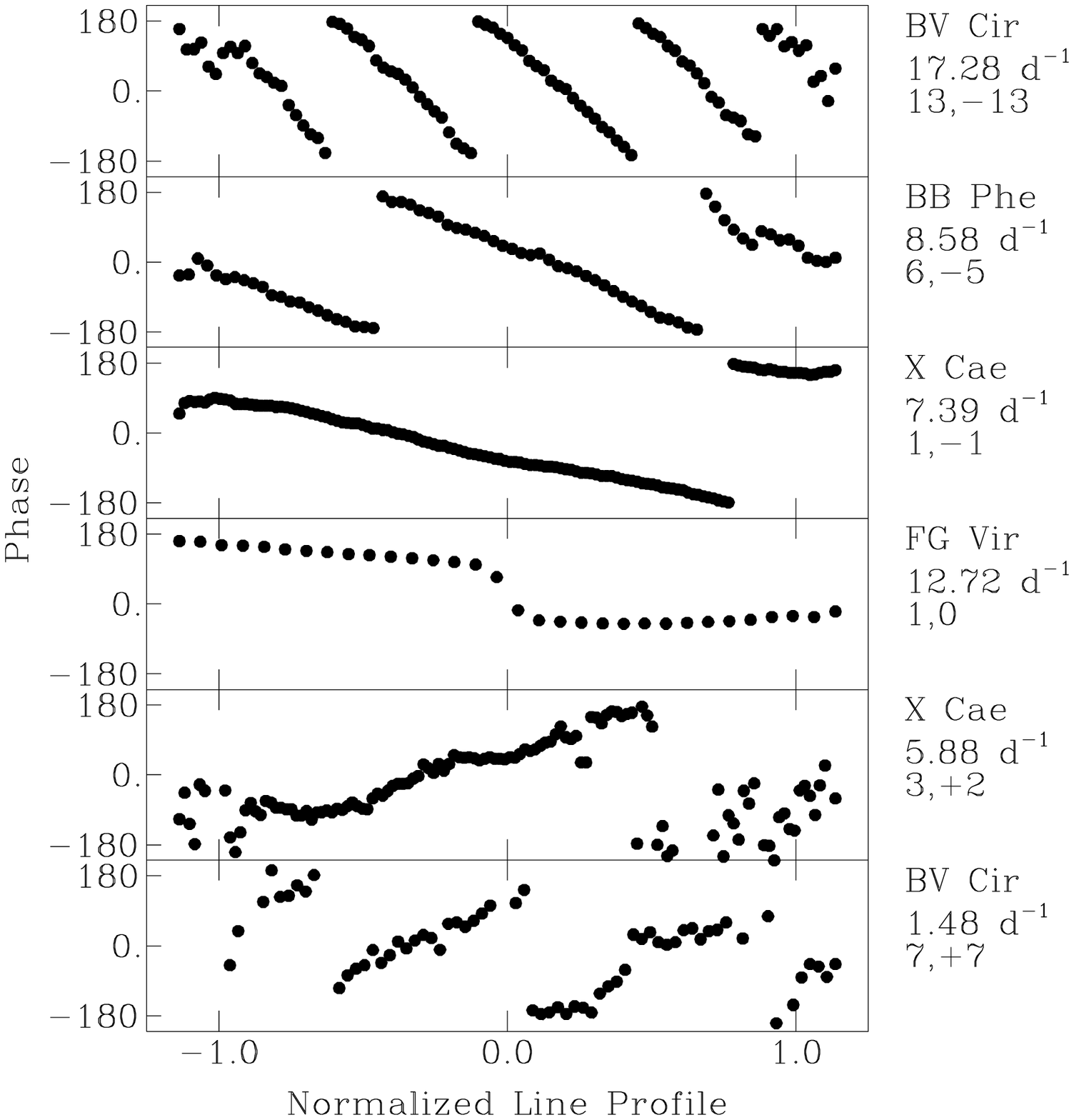}
\caption{The zoo of non-radial modes present in $\delta$ Scuti stars as
shown by the behavior of their phases across the line profile. We pass
from high--degree retrograde modes (bottom panel) through axisymmetric
modes ($m=0$, third panel from bottom) to high-degree prograde modes.
Beside each panel the name of the star to which the mode belongs is
given together with its frequency in the observer's reference frame and
the proposed $\ell$, $m$ identification.
\label{fig-11}}
\end{figure} 

\subsection{The moment time-series analysis}

The use of the line moments to study LPV was first introduced by Balona
(1986a).  Since the moments are also useful for mode identification, their 
definition is given in the paper by Aerts \& Eyer (these
proceedings). It suffices here to say that the zeroth order moment measures
the equivalent width, while the first-order one the position of the line
barycenter and hence it can be used as a measure of the stellar radial
velocity.  In his paper, Balona (1986a) gives some useful suggestions
how to derive these quantities; to this end it is also useful to look
at the paper by Balona et al. (1996).
\begin{figure}[ht]
\setlength{\textwidth}{4.3in}
\plotone{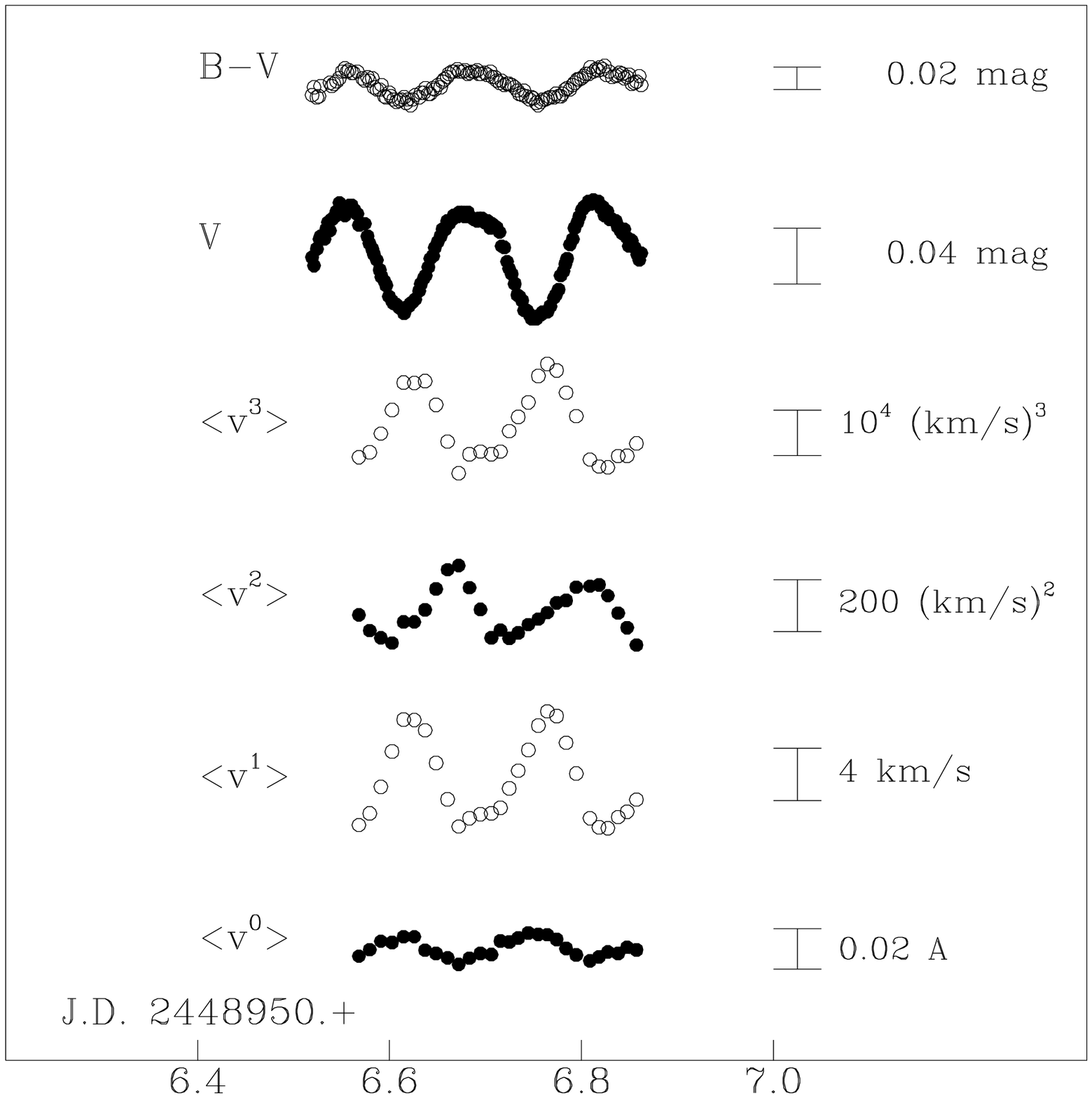}
\caption{The variations of the first moments derived from the 4501, 4508
and 4515 \AA~lines are compared with the $V$ and $B-V$ ones during one
observing night of X Caeli. All the quantities, except $V$ and $B-V$,
increase toward the top.
\label{fig-12}}
\end{figure} 
As an example Fig. 12 shows the behavior of the first three moments and
of the zero-order one during one night of observations of the star X Cae
(Mantegazza \& Poretti 1996). Their variations are compared with the
simultaneous variations in $V$ light and $B-V$ color index. It can be
observed that the first and third moment curves have a similar shape
and are in anti-phase with respect to the light variations, while the
second moment has a different behavior.

As in the case of the pixel-by-pixel time series the moment time series
can be analyzed with the same techniques used to study light-curves.
Since the moments are quantities integrated on the whole line profile,
they are more sensitive to low degree modes, because the variations due to
the high degree oscillations tend to be averaged out by the integration.

One interesting property of the moments is the fact that axisymmetric
modes contribute with their fundamental frequency only to the variations
of odd moments.  Therefore the frequency analysis of the moment time
series offers an easy way to detect these modes, and moreover, since often
we face stars with a lot of excited modes, reducing the number of the
modes affecting even moment time series makes it easier to analyze them.
As an example of this fact we show in Table 1 the results of  the
frequency analysis of the first 5 moments of FG Vir (1995 campaign,
Mantegazza et al. in preparation). The detected frequencies in each
moment time series are listed for each mode beside the well known and
reliable values derived from light-curve analysis (Breger et al. 1998).
Since the spectroscopic baseline is rather short (5 consecutive nights)
and the campaign was single site, some 1 \cd~aliases were picked up
instead of the correct values.
\begin{table}
\caption{ Frequencies of the terms detected from the analysis of the
variations of the first five moments of FG Vir compared with the accurate
frequencies detected from the light-curve analysis}
\label{tbl-1}
\begin{center}
\footnotesize
\begin{tabular}{rrrrrl}
\tableline
\\
$<v^1>$&$<v^2>$&$<v^3>$&$<v^4>$&$<v^5>$&Photom.\\
\multicolumn{6}{c}{\cd}\\
\tableline
\\
10.17 & 9.17&9.21&9.19&8.19&9.199\\
9.65 & &9.66&&9.60&9.656\\
11.16&&12.13&&12.03&12.154\\
12.72&&12.72&&12.73&12.716\\
17.11&&17.01&&&16.074\\
19.92&&19.92&&19.91&19.868\\
21.07&&21.07&&&21.052\\
21.32&&20.30&&21.28&20.288\\
20.53&&20.55&&20.60&21.551\\
23.43&22.47&&22.46&&23.403\\
23.17&&23.17&&23.19&24.200+24.228\\
34.14&&33.17&&34.16&34.119\\
\tableline
\hline
\end{tabular}
\end{center}
\end{table}
While up to 13 modes have been detected in the odd moments, the analysis
of the even ones has allowed the detection of only two.  According to what
we said above it is evident that most of the modes should be axisymmetric.
This is not surprising in view of the fact that the star is probably
seen almost pole on, and therefore the detection of axisymmetric modes
is favored by a selection effect.

\subsection{Fourier Doppler Imaging (FDI)}

This technique was first introduced to study LPV in $\delta$ Scuti
stars by Kennelly et al. (1992) and successively developed and improved
by Kennelly et al. (1998). A theoretical study of its properties has
been performed by Hao (1998).  This technique allows at the same time
the detection of the modes producing LPV and the estimation of their
non-radial degree ($\ell$).

Because it is based on the computation of a two-dimensional Fourier
Transform, which considers at the same time temporal and spatial
variations, this technique is mainly sensitive to high-degree modes (see
Kennelly et al. 1998, where these authors show also that the technique
is more sensitive to odd azimuthal order modes than to even ones) and
requires targets with rather high $v \sin i$'s.

For a description of it and an example of a two-dimensional Fourier
amplitude spectrum  see the paper by Aerts \& Eyer (these proceedings,
Fig. 8).  One of the advantages of this method is that, by separating
the detected modes both in temporal and spatial frequency, it allows,
even if the temporal baseline is not adequate, the resolution of modes
with close temporal frequencies but very different spatial ones.

\section{Recent campaigns}

Tables 2 and 3 summarize the  most recent spectroscopic campaigns on
$\delta$ Scuti stars.  In Table 2 we give for each star the epoch of the
observations, the number of sites contributing to the campaign, whether
or not there is simultaneous photometry, the adopted analysis technique
and finally the reference.  Table 3 gives more technical details such as
the resolution, the spectral range, the average $S/N$ of the individual
spectrograms at the continuum level, the number of observing nights,
the total useful observing time, the number of gathered spectra, the
average exposure time and finally the lines (or their number if they were
many) whose LPV were analyzed.  

\begin{table}
\caption{Recent campaigns to study LPV in $\delta$ Scuti stars (FDI:
Fourier Doppler Imaging; Mom: moment analysis; Pix: pixel-by-pixel
analysis.  }
\label{tbl-2}
\begin{center}
\footnotesize 
\begin{tabular}{lcrlrlr} 
\tableline 
Star&Epoch&No.&Simult.&Analysis&Reference\\
    &     &  Sites &Photom.&Technique\\
\tableline $\tau$ Peg & Oct. 90 & 1& Yes& FDI&Kennelly et al. 1992\\
           & Oct. 95 & 1&No& FDI&Kennelly et al. 1998\\
$\theta^2$ Tau & Oct. 90 & 1&No&FDI& Kennelly \& Walker 1996\\
               & Dec. 92 & 4&No&FDI& Kennelly et al. 1996\\

FG Vir & Apr. 92& 1& Yes& Mom&Mantegazza et al. 1994\\
       & Apr. 95& 1& Yes& Pix, Mom& Mantegazza et al. in preparation\\
X Cae & Nov. 92& 1&Yes& Mom, FDI &Mantegazza \& Poretti 1996\\
      & Nov. 96 & 1&Yes& Pix &Mantegazza et al. 1999\\
$\upsilon$ UMa & Apr. 93 & 1&No & FDI &Korzennik et al. 1995\\ BB Phe &
Sep. 93 &1& Yes&Pix&Bossi et al. 1998\\
       & Oct. 97 &1& No& Pix& Mantegazza \&Poretti 1999\\
HD 101158 & Apr. 94& 1&No& Pix& Mantegazza 1997\\ $\theta$ Tuc &
Sep. 96 &1&No& Pix, Mom&De Mey et al. 1998\\ BV Cir & Jun. 96& 1 & Yes&
Pix&Mantegazza et al. in preparation\\
     & Jun. 98& 1&No & Pix& Mantegazza et al. in preparation\\
V480 Tau& Nov. 96&4&No& FDI& Hao et al. 1998 \\ $\epsilon$ Cep&Sep. 97&
3&No&FDI& Kennelly et al. 1999\\ \tableline \hline \end{tabular}
\end{center} \end{table}

As we can see from Table 3 the observing parameters generally satisfy the
requirements we derived in Section 3. We note moreover that moving from
the oldest to the most recent campaigns their duration tends steadily
to increase. This fact testifies that the observers became increasingly
convinced that, given the complexity of the pulsation spectra, in order to
get reliable mode detections baselines rivaling those of the photometric
campaigns were necessary.

\addtocounter{table}{1}
  
Fig. 13 shows the position of these objects in the color-magnitude
diagram.  We can see that they are rather uniformly distributed across
the instability strip in the giant region. Main Sequence objects are
generally missing because their fainter apparent magnitudes make it
difficult to get spectrograms satisfying the observational constraints
discussed in Section 3.

\newpage

\begin{figure}[p]
\plotfiddle{tab3.ps}{15cm}{0}{100}{100}{-315}{-260}
\end{figure}

\begin{figure}[ht]
\setlength{\textwidth}{3.5in}
\plotone{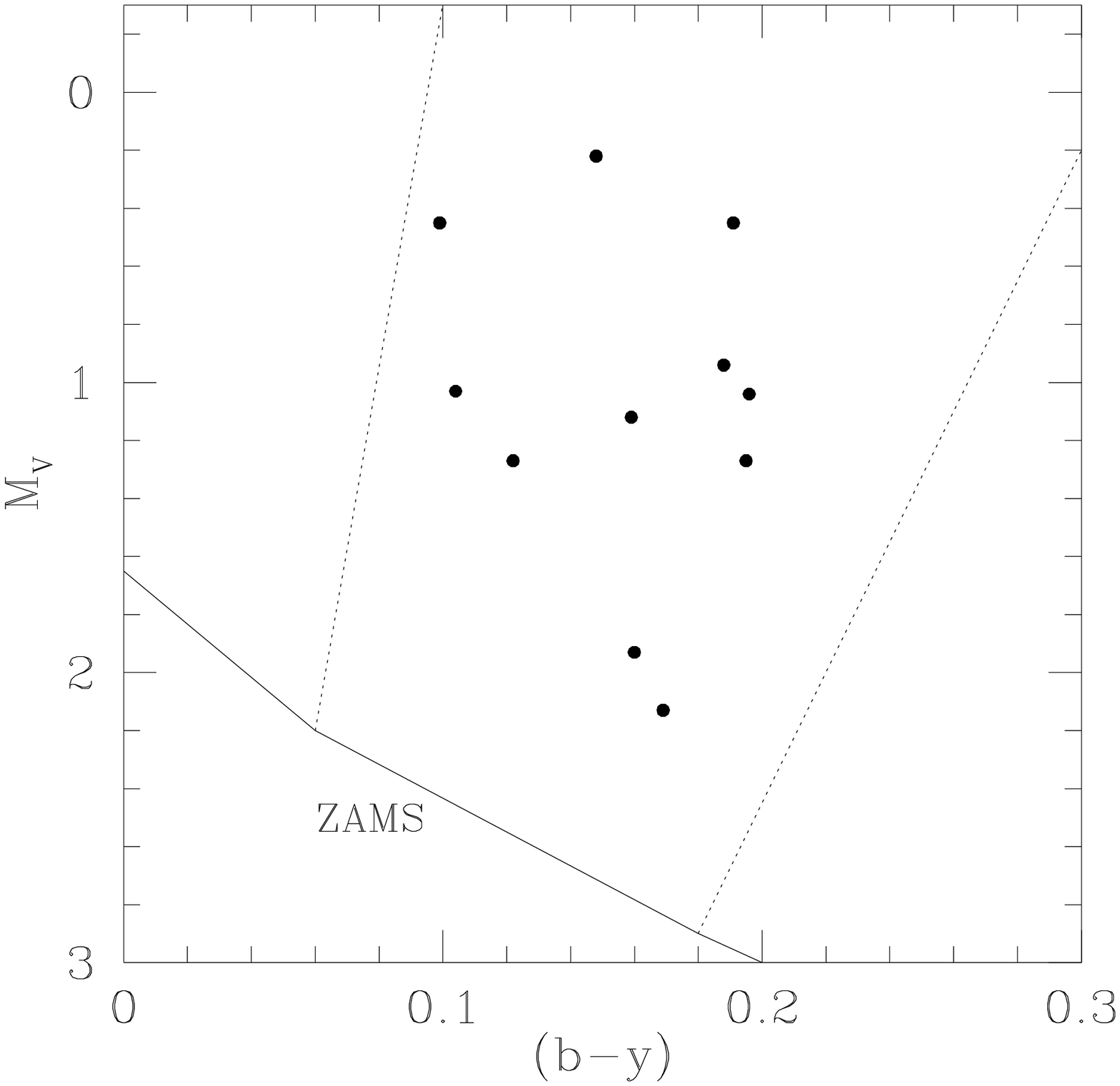}
\caption{The targets of the most recent spectroscopic campaigns in the
color-magnitude diagram. The dashed lines individuate the borders of
the instability strip.
\label{fig-13}}
\end{figure}

Table 4 summarizes the number of periods detected in each of these stars
according to the technique adopted to analyze LPV.  For comparison in the
same table  the number of periods derived from the light-curve analysis
are reported.  The uncertain detections are given within brackets.
The RV column reports the detection from the radial velocity variations,
where here for radial velocities we mean those derived with techniques
different from the first moment computation (for instance with correlation
techniques). The results regarding V480 Tau are extremely preliminary
(Hao, private communication) and many periods are expected to be found.

It is possible to see that in the cases where the temporal baseline is
adequate the number of spectroscopic detections is comparable with those
of photometric detections.

Some of the best spectroscopically studied objects have only few
photometrically detected modes, because careful photometric campaigns
have not yet been performed on these stars. A better coordination in
the selection of the spectroscopic and photometric targets should be
desirable in the future in order to get the most complete 
picture possible of the pulsation spectra.

\begin{table}
\caption{Number of periods detected with the different spectroscopic
techniques compared with the number of periods detected from light-curve
analysis. The last column gives the number of modes detected from
spectroscopic analysis but not seen in the photometric data}
\label{tbl-4}
\begin{center}
\footnotesize
\begin{tabular}{lrrlrlr}
\tableline
Star&Pix&FDI&Mom&RV&Photom. &Purely\\
&&&&&&Spectr.\\

\tableline
28 Aql &  &&(2)&2&2\\
$k^2$ Boo& 1& &&&1+(1)\\
$\gamma$ Boo & 1&&&&1\\

$\tau$ Peg & 31&&&&1+(1)&29\\
 
$\theta^2$ &&7&&5&8&6\\

FG Vir & 13&&&&24+(8)\\
      
X Cae & 14&13&&17\\
      
$\upsilon$ UMa & &3&&&1\\
BB Phe & 8&&10+(2)&&10&3\\
       
HD 101158 & 2+(3)&&1+(2)&&3\\
$\theta$ Tuc& 4&&4&&10&3\\
BV Cir & 14&&&&9&7\\
V480 Tau& &1+....&&&1\\
$\epsilon$ Cep&&$\sim$42&&&2+(1)\\
 
\tableline
\hline
\end{tabular}
\end{center}
\end{table}

Figures 14 and 15 show the best examples of pulsation spectra derived
from spectroscopic analysis. For each star the projected rotational
velocity is given below its name. The epoch of the observations is also
given at the right of the star's names because, as we will show in the next
subsection, the pulsation spectra can change considerably with the time.
The heavy vertical segment at the top of each panel marks the expected
position of the fundamental radial mode.  In the case of $\theta$ Tuc
this line is dashed, because it belongs to a binary system, so the
evaluation of its physical parameters is uncertain, and therefore the
position of this line gives only the lower limit.

We see that in the cases of FG Vir, X Cae, BV Cir and $\tau$ Peg there
are a few modes with frequencies in the observer's reference frame below
this value.  In the cases of X Cae, $\tau$ Peg and for one of the modes
of BV Cir these are retrograde $p$ modes, which in the co-rotating
reference frame have frequencies above the fundamental radial value.
For the other two modes of BV Cir the question is open, because their
identification is unclear, while the two modes of FG Vir could be $g$
modes (Breger et al. 1999).

The amplitudes of the modes indicated with solid lines were derived from
pixel-by-pixel analysis or from FDI and have been arbitrarily scaled
for each star. The modes drawn with a dashed line for BB Phe, HD 101158,
and $\theta_2$ Tau  were detected from the moment or the radial velocity
analysis, and therefore their amplitudes cannot be compared with those
of the others.

Several of the modes shown in these figures have not been photometrically
detected. The number of modes with a spectroscopic detection only is
indicated for each star in the last column of Table 4.  We see therefore
the complementarity between the two approaches and the need for both if
we want to get the whole pulsation spectrum of the star.

\begin{figure}[ht]
\setlength{\textwidth}{4.3in}
\plotone{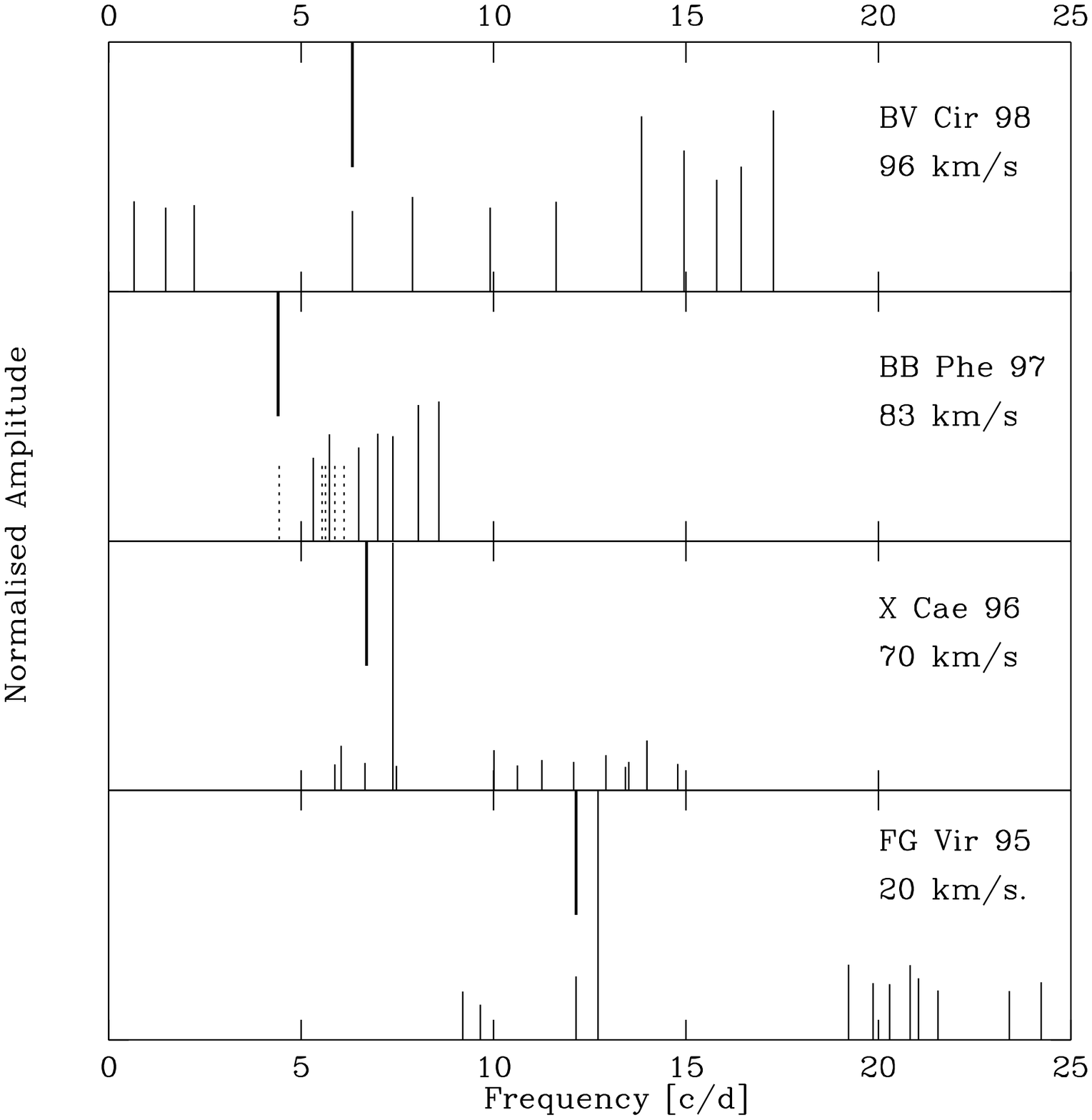}
\caption{The pulsation spectra in the observer's reference frame as
derived from spectroscopic observations of the best studied $\delta$
Scuti stars. Solid lines: modes detected from pixel-by-pixel analysis
or FDI. Dashed lines: modes detected from the line moment analysis.
Heavy lines from the top: estimated position of the fundamental radial
mode.  For each star its projected rotational velocity and the year of
the observations are reported .
\label{fig-14}}
\end{figure} 

\begin{figure}[ht]
\setlength{\textwidth}{4.3in}
\plotone{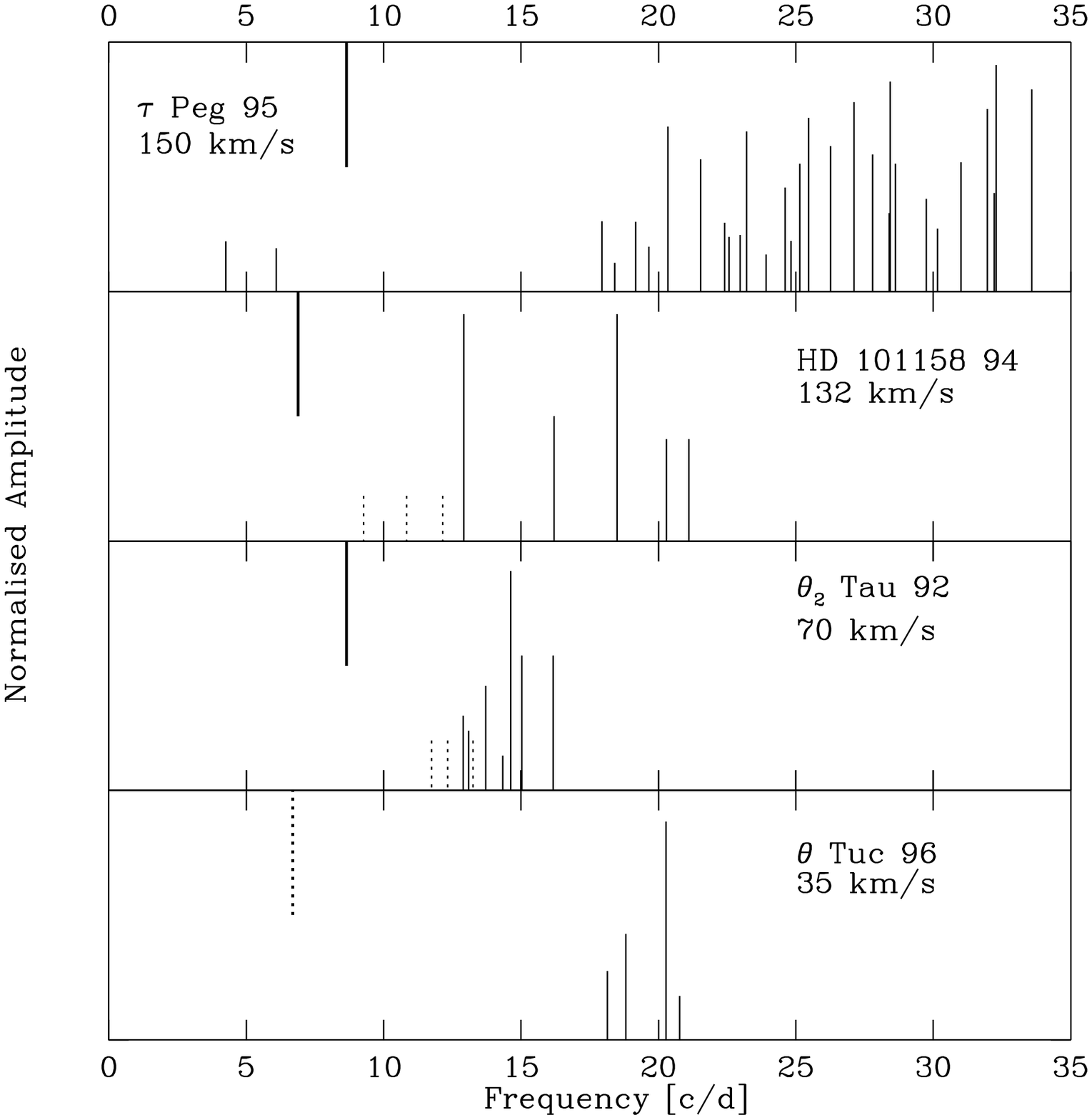}
\caption{The same as the previous figure
\label{fig-15}}
\end{figure} 

The higher the projected rotational velocity is, the higher the number
of purely spectroscopically detected modes is. This is an obvious
selection effect, because high--degree modes can be detected only in
fast rotators. This is confirmed by the fact that, for instance, for
a slow rotator such as FG Vir all the spectroscopically detected modes
also have a photometric detection.

In order to get an idea of how the pulsation spectrum of a fast rotating
$\delta$ Scuti star should appear as seen in the co-rotating reference
frame, the observed frequencies of the modes detected and identified by
Kennelly et al. (1998) with the FDI in $\tau$ Peg have been corrected
for the rotational effect assuming that all of them were sectoral. The
result is shown in the lower panel of Fig. 16.  For comparison the upper
panel shows the pulsation spectrum in the observer's reference frame and
the dashed line shows the estimated position of the fundamental radial
mode. We can see that in the co-rotating frame all the modes are clumped
in the vicinity of the fundamental radial mode.  The fact that a few
of them have frequencies slightly lower than that of the fundamental
radial one is probably due to the assumption that all the modes are
sectoral, which probably for some modes is not true, and consequently
their frequencies have been over-corrected toward low values.
\begin{figure}[ht]
\setlength{\textwidth}{4.3in}
\plotone{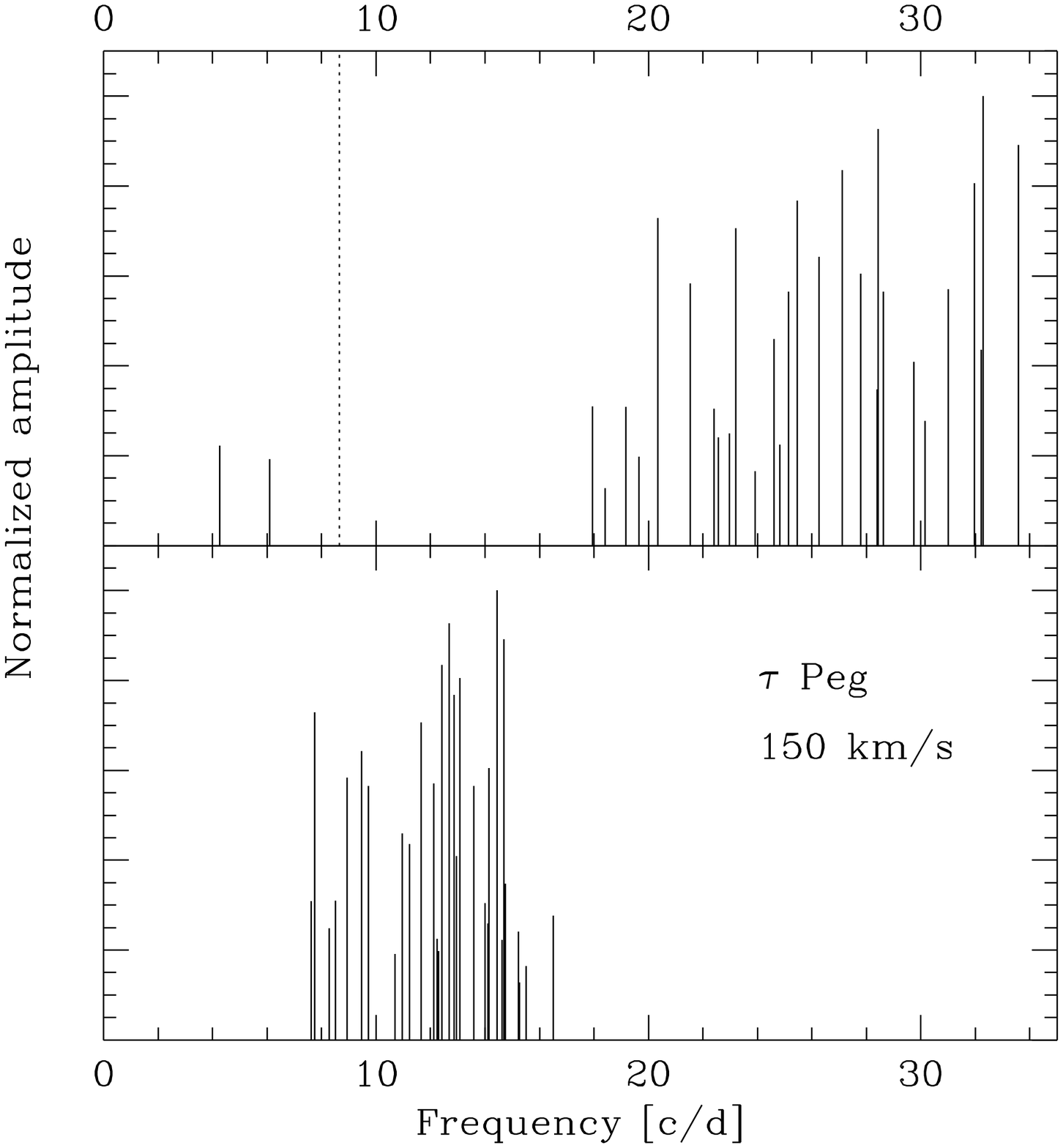}
\caption{Upper panel: pulsation spectrum of $\tau$ Peg in the observer's
reference frame; dashed line: estimated position of the fundamental radial
mode.  Lower panel: pulsation spectrum of $\tau$ Peg in the co-rotating
reference frame, assuming that all the identified modes are sectoral.
\label{fig-16}}
\end{figure} 

\subsection{Amplitude variations}

As we can see from Table 2, there are a few stars for which observations
have been performed in two campaigns. The best cases are those of X Cae,
BB Phe and BV Cir.  The availability of two independent datasets is very
useful because it allows: a) the check of the reality of the modes which
have not independently been detected from photometry, b) the detection
of variations in the mode amplitudes.  For the above-quoted three stars
most of the modes have been independently detected in the two data sets
and hence we are quite confident about their reliability.

In the cases of BB Phe and BV Cir there are large variations in the
mode amplitudes. For both stars the strongest spectroscopic mode was
different in the two seasons.
\begin{figure}[ht]
\setlength{\textwidth}{4.3in}
\plotone{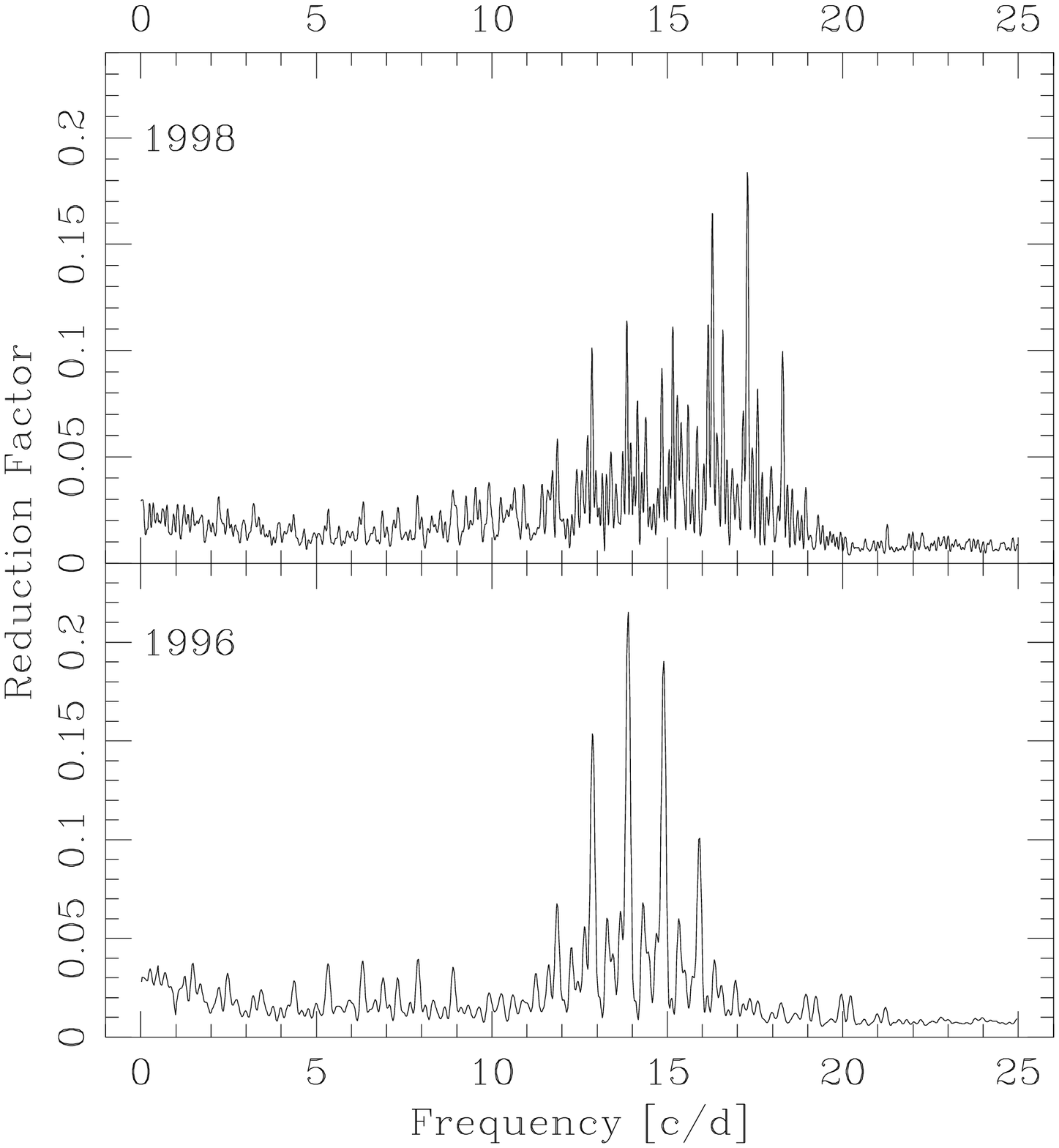}
\caption{Global least--squares power spectra without known constituents
of BV Cir in the two spectroscopic campaigns. It can be easily perceived
how the frequency content has changed.
\label{fig-17}}
\end{figure} 

\begin{figure}[ht]
\setlength{\textwidth}{4.3in}
\plotone{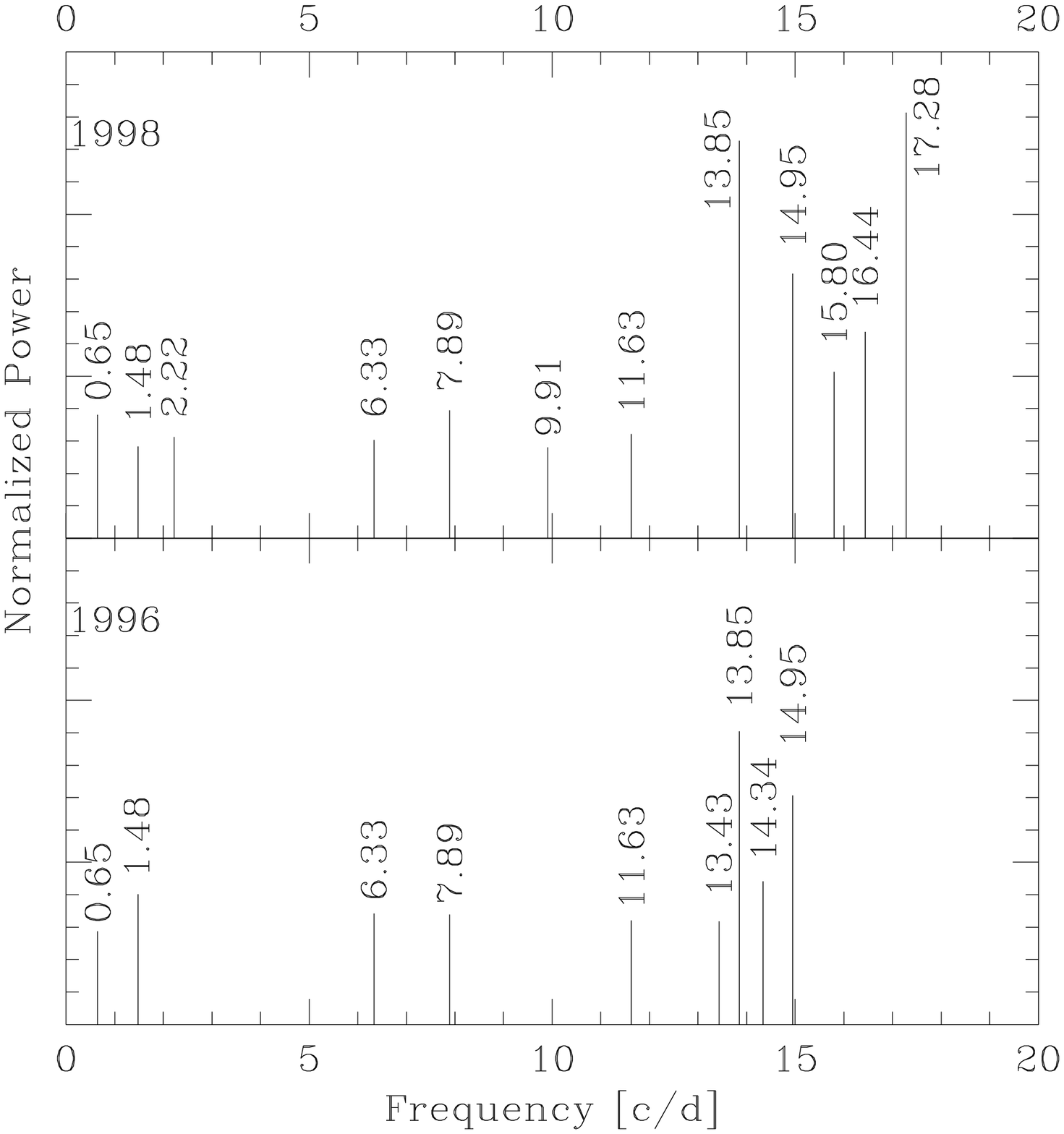}
\caption{Comparison between the pulsation spectra of BV Cir derived from
LPV in 1996 and in 1998.
\label{fig-18}}
\end{figure} 

Figure 15 shows the global least squares power spectra without known
constituents of BV Cir for the two campaigns. We clearly see the different
frequency content: in the 1996 data the strongest peak is at 13.85 \cd,
while in the 1998 data, even if this peak is still present, the highest
one is at 17.28 \cd. This last peak in 1996 was below the detectability
level. Comparing the two spectra we can also easily perceive their
different frequency resolution due to the different temporal baselines
(6 and 12 days, respectively).
\begin{figure}
\setlength{\textwidth}{4.3in}
\plotone{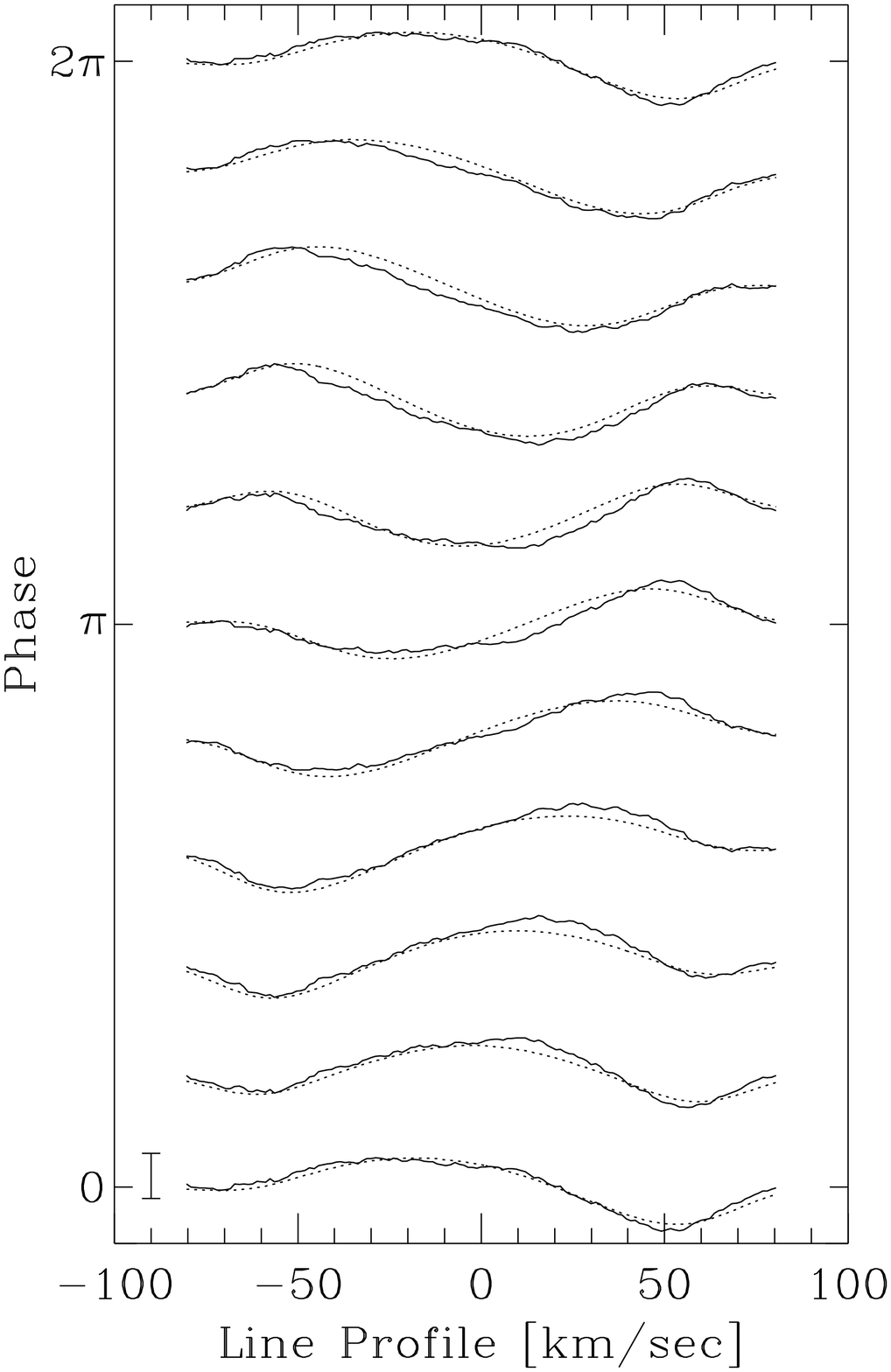}
\caption{The variations induced on the profile of the TiII 4501\AA~ line
of X Cae by the dominant mode (7.39 \cd) phase over one cycle (solid
line) and the best fitting model of the mode which gives the least
global discriminant ($\ell=1$, $m=-1$, and $i=70\deg$, dashed lines).
The small bar at the lower left indicates an amplitude of 0.02 in
continuum intensity units.
\label{fig-19}}
\end{figure} 
\begin{figure}
\setlength{\textwidth}{4.3in}
\plotone{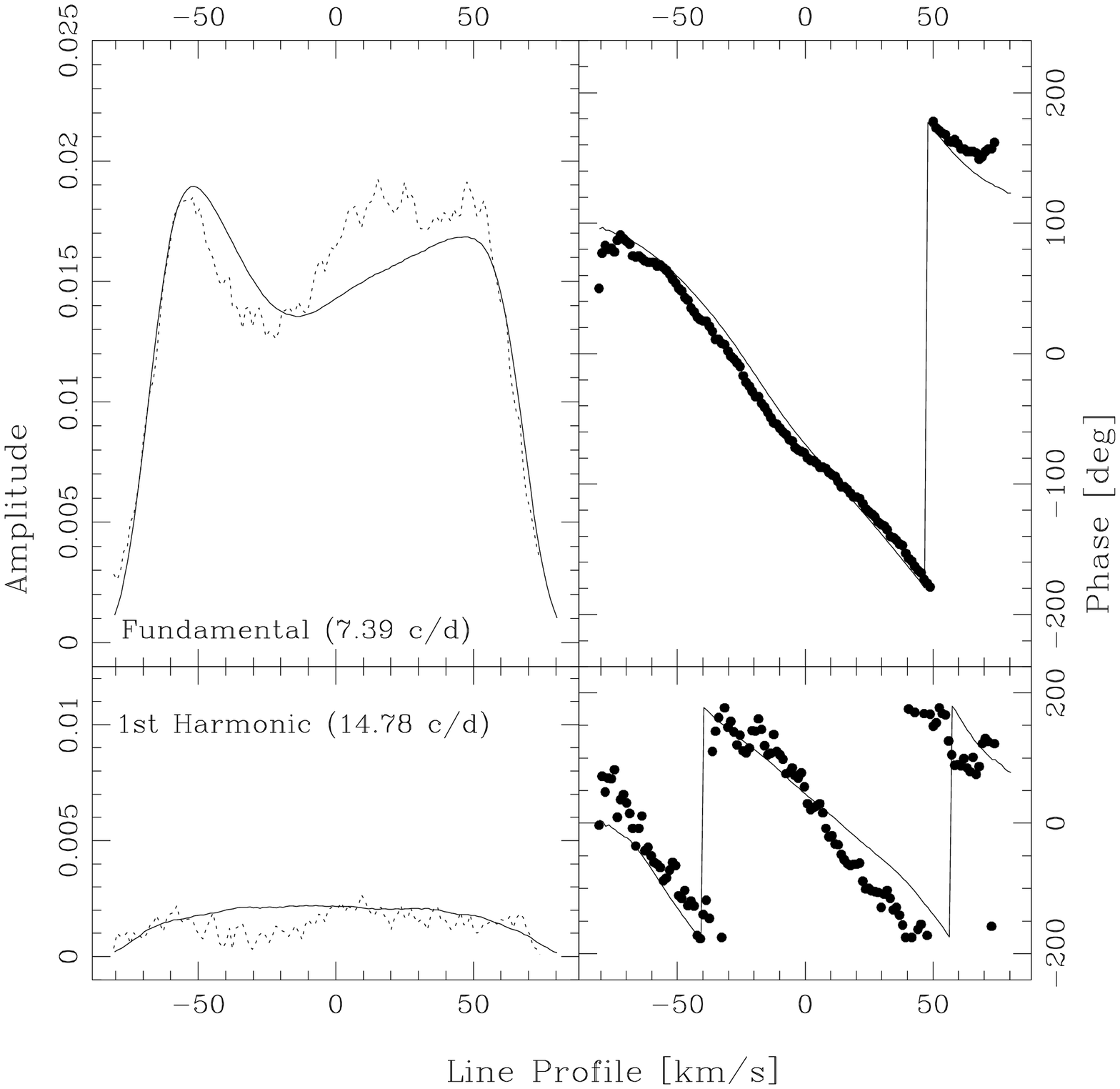}
\caption{The behavior across the profile of the 4501\AA~line of X Cae
of the model which best fit the variations induced by the dominant mode
($\ell=1$, $m=-1$ and $i=70\deg$) is compared with the same quantities as
derived from the observations.  The top panels refer to the fundamental
harmonic while the bottom one to the first harmonic.  The amplitudes
(in continuum units) are shown on the left; the phases on the right.
\label{fig-20}}
\end{figure} 

Figure 18 compares the pulsation spectrum, as derived from the
least-squares power spectrum analysis, of the two seasons.  Seven terms
are in common, but others are present only in one of the two datasets,
in particular the three highest frequency terms present in 1998, and
which are among the strongest, were not detectable in the 1995 data.

Korzennik et al. (1995) on the basis of their FDI analysis of LPV in
$\upsilon$ UMa claimed that there were strong variations in the mode
amplitudes from night to night. In this case it is easy to demonstrate
that the more plausible explanation is that this is due to beats between
unresolved modes: analyzing the data on a night by night basis, where
the observations in each night covered between 2.5 to 5 hours only,
the frequency resolution was extremely poor (between 5 to 10 \cd),
hence the expected density of the pulsation spectrum amply allows the
presence of these beats!

\section{Mode identification by LPV fit in multiperiodic stars}

In the paper by Aerts \& Eyer (these proceedings) it is stated that
the best approach to the mode identification is given by the direct
fit of LPV, but unfortunately, according to them, this is possible for
mono-periodic pulsators only. Recently, we (Bossi et al. 1998, Mantegazza
\& Poretti 1999, Mantegazza et al. 1999) have developed a technique
that allows the fit of LPV for multiperiodic pulsators and moreover, if
simultaneous spectroscopic and photometric observations are available,
the technique can proceed to the mode identification by simultaneously
fitting both variations.

In the presence of several simultaneously excited modes it is necessary
to separate the contributions of each mode to the overall LPV and light
variations.

To do this we approximate the perturbations $\Delta p_j(\lambda,t)$
induced by mode $j$ on the line profile as:
$$\Delta p_j(\lambda,t)=\sum_i A_{ij}(\lambda)\cos\left(2 \pi \,i\,\,\nu_{jt}
+\phi_{ij}(\lambda)\right)
\eqno (4)$$
where the sum is on the Fourier harmonics of the j--mode.  We can also
estimate the formal errors on $\Delta p_j(\lambda,t)$ ($\delta\Delta
p_j(\lambda,t)$) from error propagation by $\delta A_{ij}(\lambda)$
and $\delta\phi_{ij}(\lambda)$.  These last quantities as well as
$A_{ij}(\lambda)$ and $\phi_{ij}(\lambda)$ have been derived in Section
5.1 by simultaneously fitting all the detected modes and their harmonics
to the observed LPV. Usually with a $S/N$ of a few hundreds only the
fundamental harmonic of each mode is easily detectable. Only for rather
strong modes, such as for the dominant mode of X Cae (Mantegazza
et al. 1999), the first harmonic is also detectable.

Fig. 20 shows for instance the amplitude (right panels, dashed lines)
and phase (left panels, dots) variations across the 4501 \AA~ line
profile of X Cae derived from this simultaneous least-squares fit of
the fundamental frequency (upper panels) of the dominant mode and its
first harmonic (lower panels).  The reconstructed LPV (equation 4)
due to this mode are shown as solid lines in Fig.~19. In this case the
functions $\Delta p_j(\lambda,t)$ have been computed for ten equi-spaced
phases of a complete cycle.

We can try to fit the functions $\Delta p_j(\lambda,t)$ (eq. 4)
with perturbations computed with a model of a non-radial
pulsating star viewed at a certain inclination $i$.  So for
each plausible choice of $\ell,m,i$ we can build a discriminant
$$\sigma_p(\ell,m,i) = \sum_{\lambda}\sum_t{(\Delta p_j(\lambda,t)-
\Delta p_c(\lambda,t,\ell,m,i))^2\over{\delta\Delta p_j(\lambda,t)^2}}
\eqno(5)$$ where $\Delta p_c(\lambda,t,\ell,m,i)$ are the computed
profile variations which best fit the observed ones.

Moreover, if we have simultaneous photometric observations, we can obtain,
by simultaneously fitting all the terms detected in the light curve,
 amplitude and phase with respective errors for the $j$ mode. Therefore
we can calculate its light variations and relative errors ($l_j(t)$
and $\delta\l_j(t)$) and compare them with those predicted by the best
fitting models and obtain the discriminant:
 $$\sigma_l(\ell,m,i) =\sum_t(l_j(t)-
l_c(t,\ell,m,i))^2/\delta\l_j(t)^2 . \eqno(6) $$
A global discriminant is then defined as:
$$ \sigma_T(\ell,m,i)= \sigma_p(\ell,m,i)+\sigma_l(\ell,m,i). \eqno(7) $$
This is the function which is minimized with a non-linear least-squares
fit for each detected $j$ mode and for any choice of $\ell,m,i$.

The model we used to compute the synthetic line profile
variations \linebreak $\Delta p_c(\lambda,t,\ell,m,i)$ and light variations
$l_c(t,\ell,m,i)$ is the one described by Balona (1987). For each
assigned {$\ell,m,i$} the computed profile variations can be modeled
according to the amplitude and phase of vertical ($v_r$) and horizontal
($v_r$) velocities and flux variations.  For $\delta$ Scuti stars
usually $v_h \ll v_r$ and therefore in order not to introduce into the
model too many free parameters, we keep the usual theoretical relation
(e.g., Heynderickx et al. 1994) $v_h=74.4Q^2 v_r$ ($Q$ pulsation constant)
and $\psi_h=\psi_r$.

The observed light variations constrain strongly the computed flux variations,
so it is very useful to have simultaneous spectroscopic and photometric data,
otherwise in order to get meaningful physical results it is better to
fit a simplified model which considers velocity variations only, neglecting
flux variations.

By applying this approach to fit the reconstructed profile and light
variations due to the dominant mode of X Cae (solid lines of Fig. 17)
we found that the least global discriminant is supplied by a non-radial
mode with $\ell=1$, $m=-1$, and $i=70\deg$. This result is in agreement
with that derived from the independent dataset of the previous (1992)
observing season (Mantegazza \& Poretti 1998), and which was obtained
from the moment variation fits with the technique developed by Balona
(1987) (see also Balona et al. 1996), and which is different from the
approach with the moments described in Aerts \& Eyer (these proceedings).

The LPV computed according to this model are represented as dashed lines
in Fig. 19 and explains the 94\% of the variance. The same model fits
the observed $B$ variations due to this mode with a standard deviation of
 0.4 mmag.

Finally from the computed LPV we can also derive the behavior of amplitude
and phase across the line profile both for the fundamental and first
harmonic terms. These functions are represented as solid lines in the
panels of Fig. 20.

\subsection{Improved estimate of $v\sin i$ and $W_i$}

As we have said in Section 2, the average line profile tends to be wider
than the nonpulsation one and consequently there is the risk that both
the projected rotational velocity and the intrinsic line width can be
overestimated.  A better way to estimate these two quantities is to give
them as two more free parameters in the model which is non--linearly fit
to LPV.  For instance in the case of X Cae from the 4501\AA~line we get
$v\sin i$=69.0 km/s and $W_i$= 11.9 km/s from the average line profile,
while the non-linear LPV fit with the best-fitting mode ($\ell=1$,
$m=-1$) for the 7.39 \cd mode supplies $v\sin i$=68.6 km/s and $W_i$=7.7
km/s. While the estimate of the projected rotational velocity has not
appreciably changed, the intrinsic line width has been considerably
decreased.

\section{Conclusions}

In this paper we have presented the main observational characteristics
of LPV in $\delta$ Scuti stars, and we have discussed how they
can be observed and which techniques can be adopted to detect the
pulsation modes.  Moreover, a review of the main results obtained from
the application of these techniques has been given.  As we have seen,
the number of stars carefully studied is at the moment scanty, so it
is difficult to draw very general conclusions, although some facts,
which can be useful as guidelines for future studies, can be pointed out:

\begin{itemize}
\item{} The careful observation and analysis of LPV allows several 
        pulsation modes to be detected.  If data with adequate $S/N$ and
        temporal baselines are available the number of spectroscopically
        detected periods rivals the number which are photometrically
        detected.

        Many of these detections are purely spectroscopic, especially
        for stars with high $v\sin i$: photometric observations alone
        are not sufficient to get the complete pulsation spectrum.

\item{} The variability of the mode amplitudes is quite common. Sometimes
        some modes have amplitudes below the detectability
        threshold. Again, to get the complete pulsation spectrum, we
        need to observe the star at different epochs, and it may be
        desirable to get higher S/N data in order to better monitor the
        evolution of the amplitudes.

\item{} Period detections can be rather reliable: several of the
        spectroscopically detected modes have also been photometrically
        detected and moreover, for the stars with two spectroscopic
        campaigns, many of the periods have been independently detected
        in both datasets. Again, to do a good job, we need high $S/N$
        with long baselines.

\item{} The three different analysis techniques presented in Section
        4 tend to be complementary: the moment method is particularly
        suited for low-degree modes; FDI works preferably on high--degree
        ones and on stars with high $v\sin i$; the pixel-by-pixel analysis
        works both on low- and high-degree modes, and, for the detection
        of the low-degree ones, it is preferable to study stars
        with low $v\sin i$. For instance, with this technique we were
        able to detect several modes in FG Vir, which has a $v\sin i$
        of only 20 km/s.

\item{} Given the complexity of the pulsation spectra, in order to get 
        adequate frequency resolution and to avoid aliasing ambiguities,
        we need relatively long baselines (possibly comparable to the
        photometric ones) as well as multi-site campaigns.

\item{} We have already pointed out the need for data with good
        $S/N$. Given the very small amplitudes of most of the modes
        (a few thousandths of the continuum intensity), in order not to to
        merely detect them, but also to proceed to their identification,
        a $S/N$ better than 500 is desirable.

\item{} Since together with high $S/N$ data we also need high spectral and
	temporal resolutions and short exposure times (unless large
	telescopes are used for extended periods of time, which is
	almost a hopeless possibility), it is important to develop and
	to adopt techniques that add the information from many spectral
	lines. This would entail the use of medium size telescopes,
	which are more accessible.  Much work is still to be done in
	this direction, especially if the added information is used not
	merely for mode detection but for identification, too.

\item{} As an immediate consequence of the previous items, it is evident that
        the best strategy to get a complete picture of the pulsation
        behavior of a $\delta$ Scuti star is to observe it simultaneously
        both photometrically and spectroscopically for several seasons and
        in the context of multi-site campaigns. In this respect the few
        researchers active in the field should coordinate their efforts
        in order to study a few carefully selected targets.

\end{itemize}

Finally in the last part of this paper an approach to mode detection
by LPV fitting in multiperiodic variables has been presented. This
approach at the moment is not free of problems; for example, a better
model of the LPV to fit to the data is needed, a model that at the same
time is not too cumbersome, because the routine which uses it is called
thousands of times during the minimization procedure. Another problem
is the same one faced by the moment method described by Aerts \& Eyer
(this proceedings), i.e., the lack of confidence intervals for the
minima of the discriminant, which makes it difficult to compare the
competing modes.  This second problem could be rather easily overcome
once the first has been successfully addressed and the errors on the
reconstruction of the LPV produced by each mode correctly estimated,
because the discriminants (eqs. 5--7) have a $\chi^2$ shape and therefore
their statistical properties are well known.

\acknowledgments

I am grateful to E. Poretti for a critical reading of the manuscript and to
M. Bossi for some fruitful discussions.


\end{document}